\documentstyle[prl,twocolumn,aps,psfig]{revtex}

\begin{document}
\draft
\preprint{}
\title{Piezoelectricity and Piezomagnetism :\\
Duality in Two-Dimensional Checkerboards
}

\author{Leonid G. Fel}
\address{School of
Physics and Astronomy, Raymond and Beverly Sackler Faculty of Exact
Sciences\\Tel Aviv University,
Tel Aviv 69978, Israel\\ e-mail: lfel@post.tau.ac.il}

\date{\today}

\maketitle

\def\be{\begin{equation}}
\def\ee{\end{equation}}
\def\p{\prime}


\subsection*{Abstract}

The duality approach in 2-{\it dim} two-component regular checkerboards was extended onto
piezoelectricity and piezomagnetism problems. There are found a relation for effective
piezoelectric and piezomagnetic modules for the checkerboard with 
$p6^{\prime} mm^{\prime}$-plane symmetry group ({\em dichromatic triangle}).


\noindent

\widetext
\narrowtext

${\bf 1.\;\;Introduction.}$

Duality transformation in a 2-{\it dim} heterogeneous composites discovered by 
Keller \cite{keller64} and Dykhne \cite{dykhne70} has restricted number of the 
physical contexts where it could be applied. The dual symmetry is based upon the 
simple observation that any 2-{\it dim} divergence-free field when rotated locally 
at each point by $90^{o}$ becomes curl-free and vice versa. This leads to the fact
that effective physical properties of 2-{\it dim} two-component composites like an 
electric conductivity ${\widehat \sigma_{\sf ef}}$, thermal conductivity ${\widehat
\kappa_{\sf ef}}$ and other 2-nd rank symmetric tensors ${\widehat \upsilon_{\sf ef}}$
don't depend in some sense of composite's  micro-structure, being universal
\cite{mendel75}, and can be described making use of similarity of the dual problems
\begin{equation}
{\bf Y}={\widehat \upsilon}\cdot{\bf X}\;,\;\;div{\bf Y}=rot{\bf X}=0\;,\;\;
{\widehat \upsilon}_{ij}={\widehat \upsilon}_{ji}\;
\label{dykne1}
\end{equation}
with duality relation for non-isotropic structures \cite{dykhne70}, \cite{shvid83}
\begin{equation}
\det {\widehat \upsilon}_{\sf ef}=
\sqrt{\det {\widehat \upsilon}_{\sf a}\cdot \det {\widehat \upsilon}_{\sf b}}\;,
\label{dykne2}
\end{equation}
where subscripts "{\sf a}" and "{\sf b}" correspond to the {\sf a} and {\sf b}
components of composite respectively, while "{\sf ef}" denotes an effective medium.

The  further attempts \cite{Cherk92} of extension the dual symmetry onto 2-{\it dim}
elasticity
\begin{eqnarray}
{\bf u}^{ij}={\widehat {\cal K}}^{ij}_{kl}\cdot
{\bf \tau}_{kl}\;,\;\;\sum_{j}\partial_{j}{\bf \tau}_{ij}=0\;,\\
\label{milt1}
\partial^{2}_{yy} u^{xx}\;+\;\partial^{2}_{xx} u^{yy}\;=2\;\partial^{2}_{xy} u^{xy}
\label{milt2}
\end{eqnarray}
had shown an absence of duality in the sense mentioned above. The main problem on this
way makes unreducibility of an equation (\ref{milt2}) for the strain tensor 
${\bf u}^{ij}$ to the curl-free form. Therefore one can not write a duality
relation for the 4-th rank compliance tensor ${\widehat {\cal K}}^{ij}_{kl}$.

In this paper we consider the physical phenomena dealt with 3-rd rank tensors, when
a high symmetry of 2-{\it dim} two-component checkerboard makes it possible to
exploited the duality transformation. The most known are piezoelectricity and
piezomagnetism.

\vskip 0.5 cm
${\bf 2.\;\;Symmetry\;\;considerations.}$

Let us consider a homogeneous medium submitted under mechanical stress ${\widehat
\tau}$ which produces a dielectric displacement ${\bf D}$ and magnetic induction ${\bf B}$
\begin{eqnarray}
D^i={\widehat \gamma}_{jk}^i\cdot \tau_{jk}\;,\;\;
B^i={\widehat \beta}_{jk}^i\cdot \tau_{jk}\;, \nonumber \\
div\;{\bf D}=div\;{\bf B}=0\;\;,\;\;\sum_{j}\partial_{j}\tau_{ij}=0\;,
\label{cons1}
\end{eqnarray}
where the piezoelectric
${\widehat \gamma}_{jk}^i$ and piezomagnetic ${\widehat \beta}_{jk}^i$ coefficients
are a polar and an axial tensors  of the rank 3 respectively with the inner symmetry
$[V^{2}]\times V$ in Jahn notations \cite{jahn49}
\[ {\widehat \gamma}_{jk}^i={\widehat \gamma}_{kj}^i\;,\;\;
{\widehat \beta}_{jk}^i={\widehat \beta}_{kj}^i\;.\]

The piezoelectric properties in 3-{\it dim} media are possessed by the crystals 
belonging to 20 point symmetry groups, the restrictions came only from the 
crystallographic standpoint, the piezoelectricity vanishes for all medium with point 
groups contained an inversion and for cubic group $O$. The piezomagnetism 
is certainly possible only in the  crystals with the magnetic symmetry belonging 
to 90 (31 ferromagnetic and 59 anti-ferromagnetic of the 1-st type) magnetic groups 
\cite{dzyal57}. The coexistence of piezoelectric and piezomagnetic properties in
anisotropic media was discussed in \cite{alsh90}.
A wide class of materials where the piezoelectric and piezomagnetic
properties coexist is listed in \cite{chup82}. Practically all such
materials are synthetic compounds.

In 2-{\it dim} media we have a drastic decrease in a number of symmetries which 
enabled to find both piezoelectric  and piezomagnetic properties. These are 5 
ferromagnetic point groups : $C_1$, $C_s$, $C_3$, $C_s(C_1)$, $C_{3v}(C_3)$ and 
1 anti-ferromagnetic of the 1-st type group $C_{3v}$.  In general case of 
group $C_1$ there are only 6+6=12 independent piezomodules due to the inner symmetry
$[V^{2}]\times V$. In 2-{\it dim} medium it is convenient to represent each of tensors 
${\widehat \gamma}_{jk}^i$,$\;{\widehat \beta}_{jk}^i$ by two non-symmetric tensors of
the 2-nd rank
\begin{eqnarray}
{\widehat g}_{\sf 11}=
\left(\begin{array}{cc}
\gamma_1 & \gamma_3 \\
\gamma_2 & \gamma_4\end{array}\right)\;,\; \nonumber
{\widehat g}_{\sf 12}=
\left(\begin{array}{cc}
\gamma_3 & \gamma_5 \\
\gamma_4 & \gamma_6 \end{array}\right)\;,\\
{\widehat g}_{\sf 21}=
\left(\begin{array}{cc}
\beta_1 & \beta_3 \\
\beta_2 & \beta_4\end{array}\right)\;,\;
{\widehat g}_{\sf 22}=
\left(\begin{array}{cc}
\beta_3 & \beta_5 \\
\beta_4 & \beta_6 \end{array}\right)\;,
\label{matrix3}
\end{eqnarray}
where $\gamma_1=\gamma_{xx}^x,\;\gamma_2=\gamma_{xx}^y,\;\gamma_3=\gamma_{xy}^x=
\gamma_{yx}^x,\;\gamma_4=\gamma_{xy}^y=\gamma_{yx}^y,\;\gamma_5=\gamma_{yy}^x,\;
\gamma_6=\gamma_{yy}^y$. The notations for $\beta_{jk}^i$ are chosen respectively, i.e.
$\beta_1=\beta_{xx}^x$, etc.

An increase of symmetry reduces a number of piezomodules
in the following way : 

$\underline{C_{s}\;,\;C_s(C_1)\;-\;6}$
\begin{eqnarray}
{\widehat g}_{\sf 11}(C_{s})=
\left(\begin{array}{cc}
\gamma_1 & 0 \\
0 & \gamma_2\end{array}\right)\;,\; \nonumber
{\widehat g}_{\sf 12}(C_{s})=
\left(\begin{array}{cc}
0 & \gamma_3 \\
\gamma_2 & 0\end{array}\right)\;,\\
{\widehat g}_{\sf 21}(C_{s})=
\left(\begin{array}{cc}
\beta_1 & 0 \\
0 & \beta_2\end{array}\right)\;,\;
{\widehat g}_{\sf 22}(C_{s})=
\left(\begin{array}{cc}
0 & \beta_3 \\
\beta_2 & 0\end{array}\right)\;.
\label{matrixC_s}
\end{eqnarray}

$\underline{C_3\;-\;4}$
\begin{eqnarray}
{\widehat g}_{\sf 11}(C_{3})=
\left(\begin{array}{cc}
\gamma_1 & \gamma_2 \\  
\gamma_2 & -\gamma_1\end{array}\right)\;,\; \nonumber
{\widehat g}_{\sf 12}(C_{3})=
\left(\begin{array}{cc}
\gamma_2 & -\gamma_1\\
-\gamma_1 & -\gamma_2\end{array}\right)\;,\\
{\widehat g}_{\sf 21}(C_{3})=
\left(\begin{array}{cc}
\beta_1 & \beta_2 \\
\beta_2 & -\beta_1\end{array}\right)\;,\;
{\widehat g}_{\sf 22}(C_{3})=
\left(\begin{array}{cc}
\beta_2 & -\beta_1\\
-\beta_1 & -\beta_2 \end{array}\right)\;.
\label{matrixC_3}
\end{eqnarray}

$\underline{C_{3v}\;,\;C_{3v}(C_3)\;-\;2}$
\begin{eqnarray}
{\widehat g}_{\sf 11}(C_{3v})=
\left(\begin{array}{cc}
\gamma & 0\\
0 & -\gamma\end{array}\right)\;,\; \nonumber
{\widehat g}_{\sf 12}(C_{3v})=
\left(\begin{array}{cc}
0 & -\gamma\\
-\gamma & 0\end{array}\right)\;,\\
{\widehat g}_{\sf 21}(C_{3v})=
\left(\begin{array}{cc}
\beta & 0\\
0 & -\beta\end{array}\right)\;,\;
{\widehat g}_{\sf 22}(C_{3v})=
\left(\begin{array}{cc}
0 & -\beta\\
-\beta & 0\end{array}\right)\;,
\label{matrixC_3v}
\end{eqnarray}
where $\gamma, \gamma_k, \beta, \beta_k$ are real numbers.

Up to this moment we have not specified a crystallographic type of two-component composite
since according to Mendelson \cite{mendel75} a duality relation is universal there.
Nevertheless an arbitrariness of the micro-structure can make algebraic equations arisen
from the duality relations completely unsolvable, e.g. in general case two-component composite
has 12 effective piezoelectric (6) and  piezomagnetic (6) modules although the
number of the corresponding equations can be much less. Therefore it is worth to specify most 
simple cases.

\begin{figure}[h]
\psfig{figure=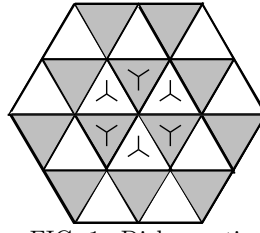,height=3cm,width=3.5cm}
\caption{Dichromatic plane mosaic $p6^{\prime} mm^{\prime}$. The principal axes of inner
medium symmetry $C_{3}$ are pointed out within the triangles.}
\label{c3v}
\end{figure}

According to the Curie principle \cite{nye64} the point symmetry group $G_{\sf ef}$ of
the physical phenomenon in composed medium is a maximal common subgroup of the 
micro-structure group $G_{\sf st}$ and the inner symmetry groups $ G_{\sf a}, G_{\sf b}$ 
of this phenomenon in both components "{\sf a}", "{\sf b}"
\begin{equation}
G_{\sf ef}=\;max\;\{\;G_{\sf st}\;\bigcap\; G_{\sf a}\;\bigcap\; G_{\sf b}\;\} \;.
\label{curi}
\end{equation}
We will look for 2-{\it dim} checkerboards with those regular {\em dichromatic}
tessellation by  polygons which are compatible with point group of the symmetry $C_3$.
From 46 {\em dichromatic plane mosaics} \cite{belov57} there is compatible only {\em
dichromatic triangle} which possesses a $p6^{\prime} mm^{\prime}$-plane group (see Fig.
\ref{c3v}). The choice of the $C_{3}$-symmetry is not accidental but concerned with the
special properties of the transport tensors ${\widehat g}_{\sf kl}$ (\ref{matrixC_3})
which we would discuss in the next section.

\vskip 0.5 cm
${\bf 3.\;\;Duality\;\;relations.}$

Let us consider a regular 2-{\it dim} two-component checkerboard with equal concentrations of the
components submitted under mechanical stress ${\widehat \tau}$ which produces a dielectric
displacement ${\bf D}$ and magnetic induction ${\bf B}$ defined in (\ref{cons1}). 
In order to make a similarity between (\ref{cons1}) and (\ref{dykne1}) more clear
we will define the following vectors
\begin{equation}
{\bf t}_{1}=\;(\tau_{xx},\tau_{xy})\;,\;\;{\bf t}_{2}=\;(\tau_{yx},\tau_{yy})
\;,\;\;div\; {\bf t}_{k}=0
\label{vect1}
\end{equation}
that makes (\ref{cons1}) more simple
\begin{equation}
{\bf D}={\widehat g}_{\sf 11}\cdot{\bf t}_1+{\widehat g}_{\sf 12}\cdot{\bf t}_2\;,\;\;
{\bf B}={\widehat g}_{\sf 21}\cdot{\bf t}_1+{\widehat g}_{\sf 22}\cdot{\bf t}_2\;.
\label{vect2}
\end{equation}
The further approach is based on the validity of the matrix identities for the
transport tensors ${\widehat g}_{\sf kl}(C_{3})$ from (\ref{matrixC_3}):
\begin{equation}
{\widehat g}_{\sf 11}(C_{3})={\widehat M}\cdot {\widehat g}_{\sf 21}(C_{3})\;,\;\;
{\widehat g}_{\sf 12}(C_{3})={\widehat M}\cdot {\widehat g}_{\sf 22}(C_{3})\;,
\label{ident1}
\end{equation}
where
\begin{equation}
{\widehat M}=\frac{1}{\beta_1^2+\beta_2^2}
\left(\begin{array}{cc}
\beta_1 \gamma_1 + \beta_2 \gamma_2 & \beta_2 \gamma_1-\beta_1 \gamma_2 \\
\beta_1 \gamma_2-\beta_2 \gamma_1 & \beta_1 \gamma_1 + \beta_2 \gamma_2\end{array}\right).
\label{ident2}
\end{equation}
This leads to unexpected relation
\begin{equation}
{\bf D}={\widehat M} \cdot {\widehat g}_{\sf 21}(C_{3})\cdot{\bf t}_1+
{\widehat M} \cdot {\widehat g}_{\sf 22}(C_{3})\cdot{\bf t}_2={\widehat M} \cdot
{\bf B}\;.
\label{ident3}
\end{equation}
The next transformation will bring a latter equation (\ref{ident3}) in the completely
similar to (\ref{dykne1}) relation between divergence-free field ${\bf D}$
and curl-free field ${\widehat {\cal R}} {\bf B}$
\begin{eqnarray}
{\bf D}&=&{\widehat M} \cdot {\widehat {\cal R}}^{-1}\cdot 
{\widehat {\cal R}} {\bf B}\;,\;\label{ident4} \\
div\;{\bf D}&=&rot{\;\widehat {\cal R}} {\bf B}=0\;,\;\;\;
{\widehat {\cal R}}=\left(\begin{array}{cc}
0 & -1 \\
1 & 0\end{array}\right)\;,\nonumber
\end{eqnarray}
where ${\widehat {\cal R}}$ is a rotation by $90^{o}$ operator 
and ${\widehat L}={\widehat M} \cdot {\widehat {\cal R}}^{-1}$ is an anti-symmetric
positively defined matrix
\[ {\widehat L}=\frac{1}{\beta_1^2+\beta_2^2}
\left(\begin{array}{cc}
\beta_1 \gamma_2-\beta_2 \gamma_1 & \beta_1 \gamma_1 + \beta_2 \gamma_2 \\
-(\beta_1 \gamma_1 + \beta_2 \gamma_2) & \beta_1 \gamma_2-\beta_2 \gamma_1
\end{array}\right).\]
Applying now the duality transformation to (\ref{ident4}) we will make
use of duality relation (\ref{dykne2}) 
\begin{equation}
\det {\widehat L}_{\sf ef}=
\sqrt{\det {\widehat L}_{\sf a}\cdot \det {\widehat L}_{\sf b}}\;.
\label{ident5}
\end{equation}
After simple algebra we obtain finally
\begin{equation} 
\frac{\gamma_{1{\sf ef}}^2+\gamma_{2{\sf ef}}^2}{\beta_{1{\sf ef}}^2+\beta_{2{\sf
e}}^2}=
\sqrt{\frac{\gamma_{1{\sf a}}^2+\gamma_{2{\sf a}}^2}
{\beta_{1{\sf a}}^2+\beta_{2{\sf a}}^2}\cdot 
\frac{\gamma_{1{\sf b}}^2+\gamma_{2{\sf b}}^2}
{\beta_{1{\sf b}}^2+\beta_{2{\sf b}}^2}}\;.
\label{ident6}
\end{equation}
If the inner symmetry  $G_{\sf a}, G_{\sf b}$ increases up to the $C_{3v}$ (\ref{matrixC_3v})
the duality relation (\ref{ident5}) looks more simple
\begin{equation}
\frac{\gamma_{\sf ef}^2}{\beta_{\sf ef}^2}=|\frac{\gamma_{\sf a}\;\gamma_{\sf b}}
{\beta_{\sf a}\;\beta_{\sf b}}|\;.
\label{ident7}
\end{equation}

Note that the equations (\ref{vect2}) do not permit to write the duality relations  
for the piezoelectric ${\widehat \gamma}_{jk}^i$ and piezomagnetic ${\widehat \beta}_{jk}^i$ 
tensors separately.

\vskip 0.5 cm
${\bf 4.\;\;Conclusion.}$ 

In the present paper we have considered the coexisting simultaneously piezoelectric and
piezomagnetic phenomena in 2-{\it dim} two-component composites with triangular
 tessellation of the plane ($p6^{\prime} mm^{\prime}$-plane symmetry group).
The  duality approach in 2-{\it dim} two-component regular checkerboards was extended onto
physical problem dealt with 3-rd rank tensors.

In the conclusion we will name some compounds which have a trigonal symmetry and
where the piezoelectricity and the piezomagnetism coexist \cite{chup82}. These are the
rare-earth maganites having the overall formula 
{\it RMnO$_{3}$}, where {\it R=Y, Ho, Er, Tm, Yb, Lu}, or {\it Sc}.
The {\it Mn} atoms lie inside the bipyramidial bonds, while the rare-earth atoms lie
inside the bipyramids. The trigonal structure in these compounds arises from the
smallness of the ionic radii of the rare-earth ions and the presence of covalent {\it
Mn-O} bonds.

\vskip 0.5 cm 
${\bf 5.\;\;Acknowledgement.}$

It is a pleasure to thank D.J.Bergman for a discussion which stimulated this work. 

This research was supported in part by grants from the Tel Aviv University Research
Authority and the Gileadi Fellowship program of the Ministry of Absorption
of the State of Israel.

\newpage



\end{document}